\documentclass{article}
\usepackage[letterpaper]{geometry}
\usepackage{spconf,graphics,hyperref}

\usepackage{wrapfig}
\usepackage{lscape}
\usepackage{rotating}
\usepackage{graphicx}
\usepackage{epstopdf}
\usepackage{caption}
\usepackage{subfigure}
\usepackage{amsmath}

\usepackage{gensymb}
\usepackage{relsize}
\usepackage{mathtools}
\DeclarePairedDelimiterX{\norm}[1]{\lVert}{\rVert}{#1}
\DeclarePairedDelimiter{\ceil}{\lceil}{\rceil}
\usepackage{hyperref}
\usepackage{tabularx}
\usepackage{listings}
\usepackage{color}
\usepackage{float}
\usepackage{csquotes}
\usepackage{tikz}
\usepackage{rotating}
\usepackage{algorithmicx}
\usepackage{algorithm}
\usepackage{pdfpages}
\usepackage{multirow}
\usepackage[noend]{algpseudocode}
\definecolor{codegreen}{rgb}{0,0.6,0}
\definecolor{codegray}{rgb}{0.5,0.5,0.5}
\definecolor{codepurple}{rgb}{0.58,0,0.82}
\definecolor{backcolour}{rgb}{0.99,0.99,0.99}
\title{MODELLING QUASI-ORTHOGRAPHIC CAPTURES FOR SURFACE IMAGING}

\name{Maniratnam Mandal  \ and \  Venkatesh K. Subramanian\thanks{Thanks to Computer Vision Lab, IIT Kanpur.}}
\address{Department of Electrical Engineering, IIT Kanpur}

\begin{document}
\maketitle

\begin{abstract}
  Surveillance and surveying are two important applications of empirical research. A major part of terrain modelling is supported by photographic surveys which are used for capturing expansive natural surfaces using a wide range of sensors- visual, infrared, ultrasonic, radio, etc. A natural surface is non-smooth, unpredictable and fast-varying, and it is difficult to capture all features and reconstruct them accurately. An orthographic image of a surface provides a detailed holistic view capturing its relevant features. In a perfect orthographic reconstruction, images must be captured normal to each point on the surface which is practically impossible. In this paper, a detailed analysis of the constraints on imaging distance is also provided. A novel method is formulated to determine an approximate orthographic region on a surface surrounding the point of focus and additionally, some methods for approximating the orthographic boundary for faster computation is also proposed. The approximation methods have been compared in terms of computational efficiency and accuracy.    
\end{abstract}


\section{Introduction}
\label{sec:intro}
Visual representation of topographical data or Digital Terrain Modelling has a wide-spread application, be it Google Maps, Navigation, Geological surveys, agriculture, disaster management, Astronomical studies and mapping of extra terrestrial surfaces, Archaeological studies, Sociological studies or even deciding national policies. 
The capturing technology has progressed a lot over the past decade and nowadays a mixture of visual, radio, infrared, laser and radar sensors are used to capture terrain information \cite{terrainmap2001}. However, natural terrains are highly uneven and for vast surfaces, when very small features are the focus of studies, it is easy to lose them while reconstruction or even while capturing. An orthographic projection or image of a surface is a holistic representation of all its features. To construct an idealistic orthographic projection of the whole surface, captures at all points need to be taken separately. This approach is highly impractical, implausible and incomprehensible due to resource limitations. Thus an approximation of local orthographic region needs to be formulated for practical purposes and methods to determine capture points for the coverage of the surface is also needed.


The physical parameters of an imaging system is guided by the optics and the sensors of the device\cite{imgparam}, among which, \textit{Field of View (FOV)} and \textit{Working Distance (WD)} are important parameters to be considered for orthographic imaging. Usually, an \textit{object-space telecentric lens}\cite{telelens2} with the entrance pupil at infinty, is used for eliminating the perception of depth and creating orthographic images. A more commonly used technology is \textit{orthophotography}. Orthophotographs are commonly used in \textit{geographic information systems (GIS)} as it is 'map-accurate'. A \textit{digital elevation model (DEM)} is often required to create accurate orthophotos. An aerial image or a satellite image of a terrain or surface which is geometrically corrected or \textit{orthorectified} such that the photo or image is essentially an orthographic projection of the terrain. An orthophotograph can be used to measure distances accurately because it is an almost accurate depiction of the Earth's surface being adjusted for topographic relief, perspective distortion and camera tilt.\cite{orthograph1}

In this paper, a mathematical approach is taken to generate orthographic captures of surfaces. In section \ref{sec:imaging_surf}, the derivation and analysis of \textit{imaging surfaces} is given and the mathematical bounds on working distance has been formulated. Section \ref{sec:ortho_appx} deals with the formulation of approximate orthography and in section \ref{sec:implement} the algorithms for computation of orthographic bounds for curves and regions for surfaces have been provided. In section \ref{sec:appx_boundary}, several methods for approximating orthographic boundary for faster computation are proposed and compared and in section \ref{sec:conclusion}, the contributions of this paper have been summarized along with potential future extensions.

\begin{figure}[H]
\centering
\subfigure[Gray-scale DEM]{\label{fig:sub1}\includegraphics[width=0.3\linewidth]{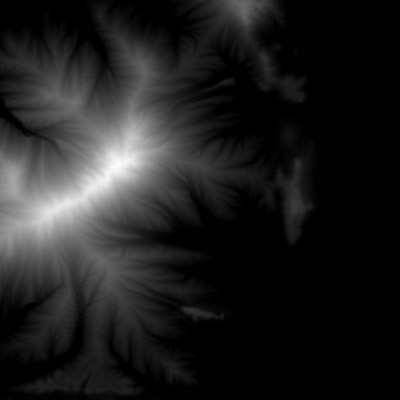}}
\hfill
\subfigure[Generated surface plot]{\label{fig:sub2}\includegraphics[width=0.4\linewidth]{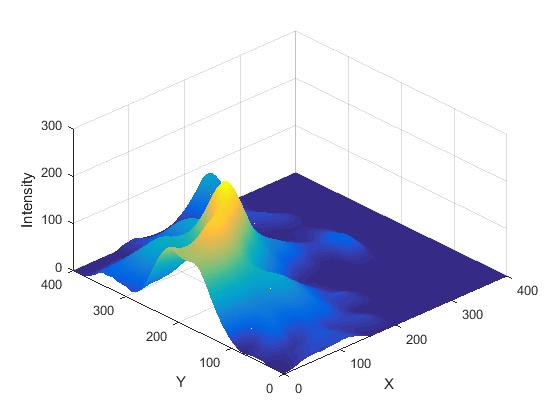}}
\caption{\textbf{a.} A gray-scale DEM denoting a terrain. (Source: Creating Heightfields and Details on Terrain RAW Files, wiki.secondlife.com). \textbf{b.} Surface plot generated in \textsc{Matlab}.}
\label{fig:3}
\end{figure}

\section{IMAGING SURFACE}
\label{sec:imaging_surf}
\subsection{Generating Surface Topographies}\label{sec:surf_topo}
A digital terrain elevation map (DTEM) is a digital image of a terrain or a topography map where the pixel intensity at a point gives the relative elevation of the point. There are various ways of allocating pixel values (Gray-scale or RGB colormap) in a DTEM. The images that were used for the purpose of this thesis are Gray-scale DTEMs, where the elevation of a point or location in the digital map can range from $0$ to $255$, the white intensity pixels denoting the highest elevation points and the black intensity pixels denoting the lowest intensity points. Fig. \ref{fig:sub1} is an example of a gray-scale digital elevation map. 

The obtained image of the height-map is first smoothed out because rough surfaces with abrupt changes creates difficulty in further processing. The double precision matrix $I$ is used as a topographical surface for later processing. The surface generated in \textsc{Matlab} for the DTEM in figure \ref{fig:sub1} is shown in figure \ref{fig:sub2}.



\subsection{Formulation}\label{sec:deriv_surf}
\begin{figure}[hbt!]
\centering

\subfigure[Surface normal plot]{\label{fig:sub3}\includegraphics[width=0.4\linewidth]{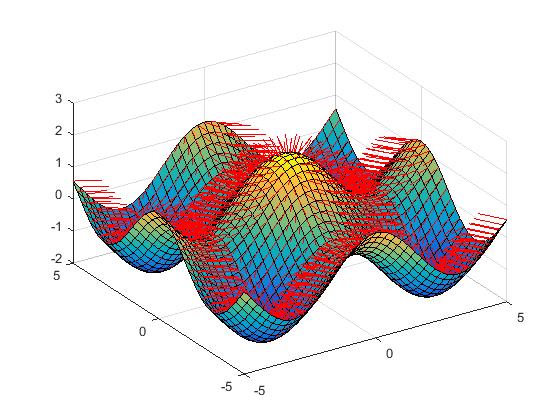}}
\hfill
\subfigure[Contour plot]{\label{fig:sub4}\includegraphics[width=0.4\linewidth]{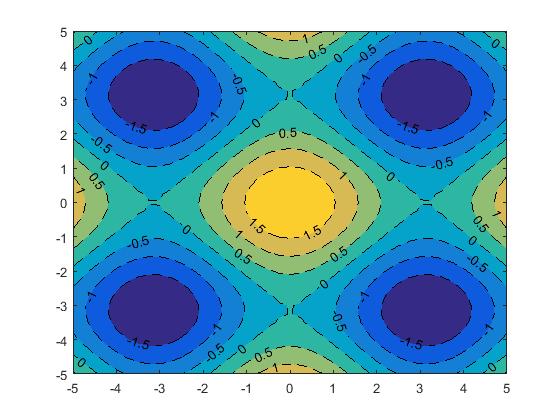}}


\subfigure[$d = 0.5$]{\label{fig:sub11}\includegraphics[width=0.4\linewidth]{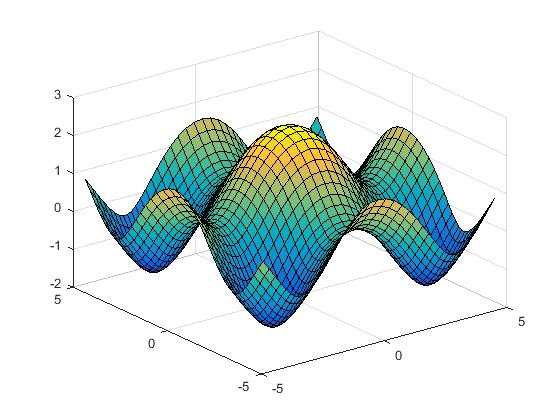}}
\hfill
\subfigure[$d = 0.5$]{\label{fig:sub12}\includegraphics[width=0.4\linewidth]{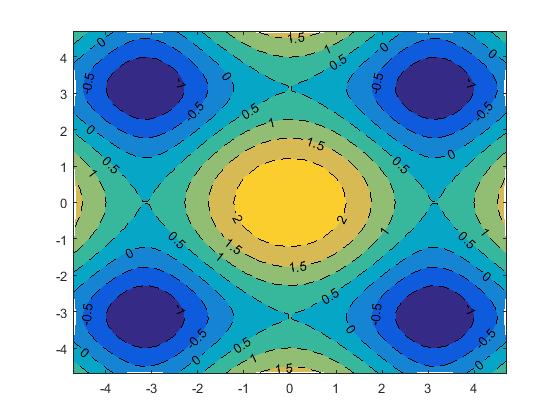}}

\subfigure[$d = 1$]{\label{fig:sub7}\includegraphics[width=0.4\linewidth]{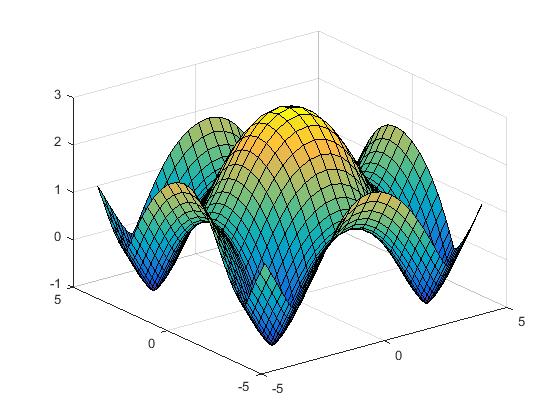}}
\hfill
\subfigure[$d = 1$]{\label{fig:sub8}\includegraphics[width=0.4\linewidth]{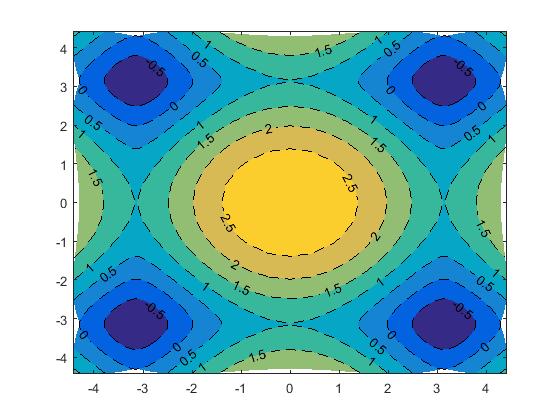}}
\caption{The surface $S$ given by $f(x,y) = cos(x) + cos(y)$ and the imaging surfaces $S'$ plotted in \textsc{Matlab}}.
\label{fig:4}
\vspace{-3mm}
\end{figure}

Let the working distance be denoted as $d$, i.e., it is assumed that to capture a point $P(x,y,z)$ on the surface $S$ (given by the bi-variate function $f(x,y)$), the camera needs to be placed at $P'(x',y',z')$ at a height $d$ along the normal to the surface at $P$. So,
\begin{equation}
    z = f(x,y).
\end{equation}
Then the surface normal at point $P(x,y,z)$ is given as
\begin{equation}\label{eq2}
\Vec{n} = \left[\frac{\partial f(x, y)}{\partial x}, \frac{\partial f(x, y)}{\partial y},-1\right]. 
\end{equation}
If variables $p$ and $q$ are defined as 
\begin{equation}\label{eq3}
    p = \frac{\partial f(x, y)}{\partial x} \ \ and \ \ 
    q = \frac{\partial f(x, y)}{\partial y},
\end{equation}
then the surface normal can be written as $[p, q, -1]$. The unit normal vector at $P$ is 

\begin{equation}\label{eq4}
\begin{aligned}
    \hat{n} & = \frac{\Vec{n}}{|{\Vec{n}}|}  = \frac{\Vec{n}}{\sqrt{p^2 + q^2 + 1}}\\
    & = \left[\frac{p}{\sqrt{p^2 + q^2 + 1}}, \frac{q}{\sqrt{p^2 + q^2 + 1}}, \frac{-1}{\sqrt{p^2 + q^2 + 1}}\right].
\end{aligned}
\end{equation}
Now, using Eq. \ref{eq4} as derived above, the corresponding imaging point, $P'(x',y',z')$, at a height $d$ from point $P$ and along the unit surface normal $\hat{n}$ is given by
\begin{equation}\label{eq5}
\begin{aligned}
    \Vec{P'} & = \Vec{P} + d\cdot\hat{n}\\
             & = \Vec{P} + \left[\frac{d\cdot p}{\sqrt{p^2 + q^2 + 1}}, \frac{d\cdot q}{\sqrt{p^2 + q^2 + 1}}, \frac{-d}{\sqrt{p^2 + q^2 + 1}}\right].
\end{aligned}
\end{equation}
Therefore the co-ordinates of point $P'(x',y',z')$ can be derived using Eq. \ref{eq5}.
The imaging surface $S'$ at imaging distance $d$ can be parameterized in terms of $x$ and $y$ as shown in Eq. \ref{eq6}.  Here $p = \frac{\partial f(x, y)}{\partial x}$ and $q = \frac{\partial f(x, y)}{\partial y}$.
\begin{equation}\label{eq6}
    \Vec{S'} = \left[
    \begin{array}{c}
         {x + \frac{d\cdot p}{\sqrt{p^2 + q^2 + 1}}} \\
         {y + \frac{d\cdot q}{\sqrt{p^2 + q^2 + 1}}} \\
         {f(x,y) - \frac{d}{\sqrt{p^2 + q^2 + 1}}}
    \end{array} \right]
\end{equation}
The surface plots in Fig. \ref{fig:4} demonstrate the imaging surfaces for the surface $S$ given by $f(x,y) = cos(x) + cos(y)$ in the range $(-5\leq x \leq 5)$ and $(-5 \leq y \leq 5)$ calculated and plotted at different values of $d$.

In case the surface $S$ is represented a double precision matrix ($I$) as shown in section \ref{sec:surf_topo}, then instead of calculating mathematical gradients ($\frac{\partial f(x, y)}{\partial x}$ and $\frac{\partial f(x, y)}{\partial y}$) for finding surface normals, the numerical gradients can be calculated as an approximation.


\subsection{Analysis of Imaging Curves}
If at any point on $S'$ is below the surface $S$, then those points are inaccessible and hence cannot be used as imaging points, i.e. if $z' <  f(x',y')$, $P'(x',y',z')$ cannot be an imaging point for $P(x,y,z)$ at height $d$. Visualizing the variation of the \textit{imaging surface} with imaging height $d$ is difficult for bi-variate functions. Hence the following analysis is done for imaging curves and virtual bounds on $d$ is derived.

Let $C$ be the target curve given by $y = f(x)$. The curve can be parametrized by vector $\Vec{P}$ as
\begin{equation}
    \Vec{P}(x) = \left[ \begin{array}{c}{x}\\{y}
    \end{array}\right] = \left[ \begin{array}{c}{x}\\{f(x)}
    \end{array}\right].
\end{equation}
Proceeding similarly as in section \ref{sec:deriv_surf}, the coordinates($x'$ and $y'$) of imaging curve $C'$ located at distance $d$ can be parametrized in terms of $x$ (ref Eq.\ref{eq10} ).
\begin{equation}\label{eq10}
    \begin{aligned}
        x' & = x - \frac{d\cdot f'(x)}{\sqrt{1 + (f'(x))^{2}}} \\ \text{and} \ \ \
        y' & = f(x) + \frac{d}{\sqrt{1 + (f'(x))^{2}}}.
    \end{aligned}
\end{equation}

If a point $P'(x',y')$ on curve $C'$ is such that it satisfies $y'(x) < f(x')$, then it lies below the curve $C$($f(x)$) and hence it cannot be accepted as a valid imaging point. In other terms for a $d$ to be valid, curves $C$ and $C'$ should not intersect at any point. This gives a mathematical bound for imaging height $d$ -
\begin{itemize}
    \item $d > 0$
    \item $d < D$ such that $ \forall \ d \geq D, \ \exists$ some $x$ in $dom(f)$,  s.t. $y'(x) < f(x')$, where $x'$ and $y'$ are as given in Eq.\ref{eq10}.
\end{itemize}

The mathematical upper bound $D$ depends on the curvature or nature of the function $f(x)$ and also the imaging range, i.e., the range of values of $x$ that is to be imaged. D can be calculated numerically by solving Eq.\ref{eq11} and applying the \textit{bisection algorithm}.
\begin{equation}\label{eq11}
\begin{aligned}
    & y'(x) = f(x') \\ 
    or, \ & f(x') =  f(x) + \frac{d}{\sqrt{1 + (f'(x))^{2}}}
\end{aligned}
\end{equation}
For $d < D$, the above equation will have no solution and for $d \geq D$, the above equation will have one or more solution(s). 

\begin{algorithm}[H]
\scriptsize
\caption{Finding upper bound $D$ by Bisection }\label{alg:bisecD}
\begin{algorithmic}[1]
\State Set lower bound of $d$, $L = 0$.
\State Set a very large upper bound $U$ such that by replacing $d = U$ in Eq.\ref{eq11}, it has a solution.
\State Set $D = (L+U)/2$.
\State Replace $d = D$ in Eq.\ref{eq11} and look for a solution.
    \If{solution exists}
        \State Set $U = D$
    \Else{
        \State Set $L = D$
    }

\State Repeat steps 5-8 until convergence criterion is met.
\EndIf
\end{algorithmic}
\end{algorithm}
For some smooth functions, there may not be any upper limit on $d$ (i.e. $D = \infty$). For those functions, Eq. \ref{eq11} does not have a solution for any $d > 0$, i.e. $y'(x) > f(x')$ for all positive values of $d$ and all $x$ in $dom(f)$. However, the practical upper bound depends on the limitations of resolution of the capturing device and also the concerned application. 

\begin{figure}[H]
\centering

\subfigure[$f(x) = sin(x)$]{\label{fig:sub11}\includegraphics[width=0.45\linewidth]{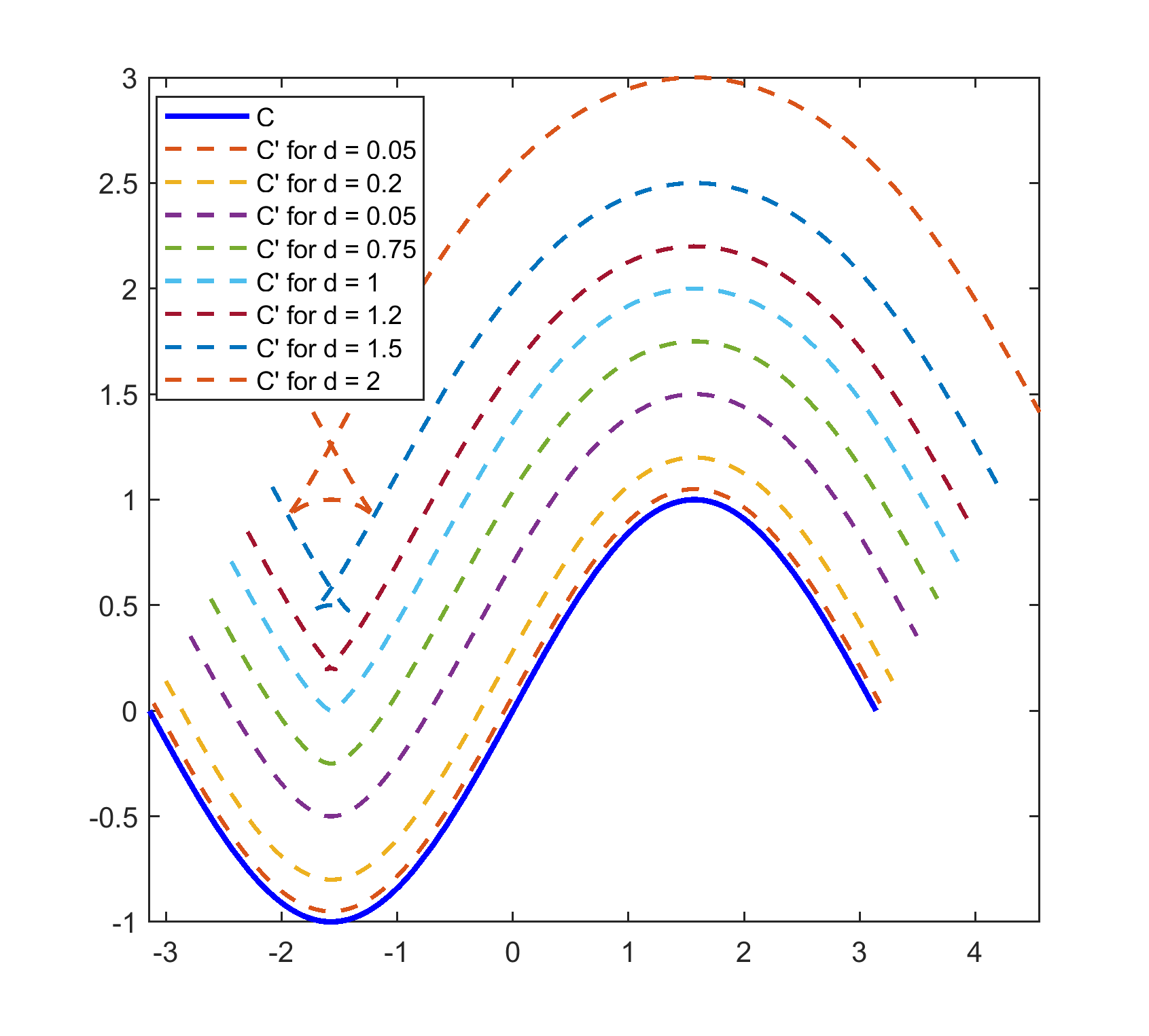}}
\hfill
\subfigure[$f(x) = x^2$]{\label{fig:sub12}\includegraphics[width=0.45\linewidth]{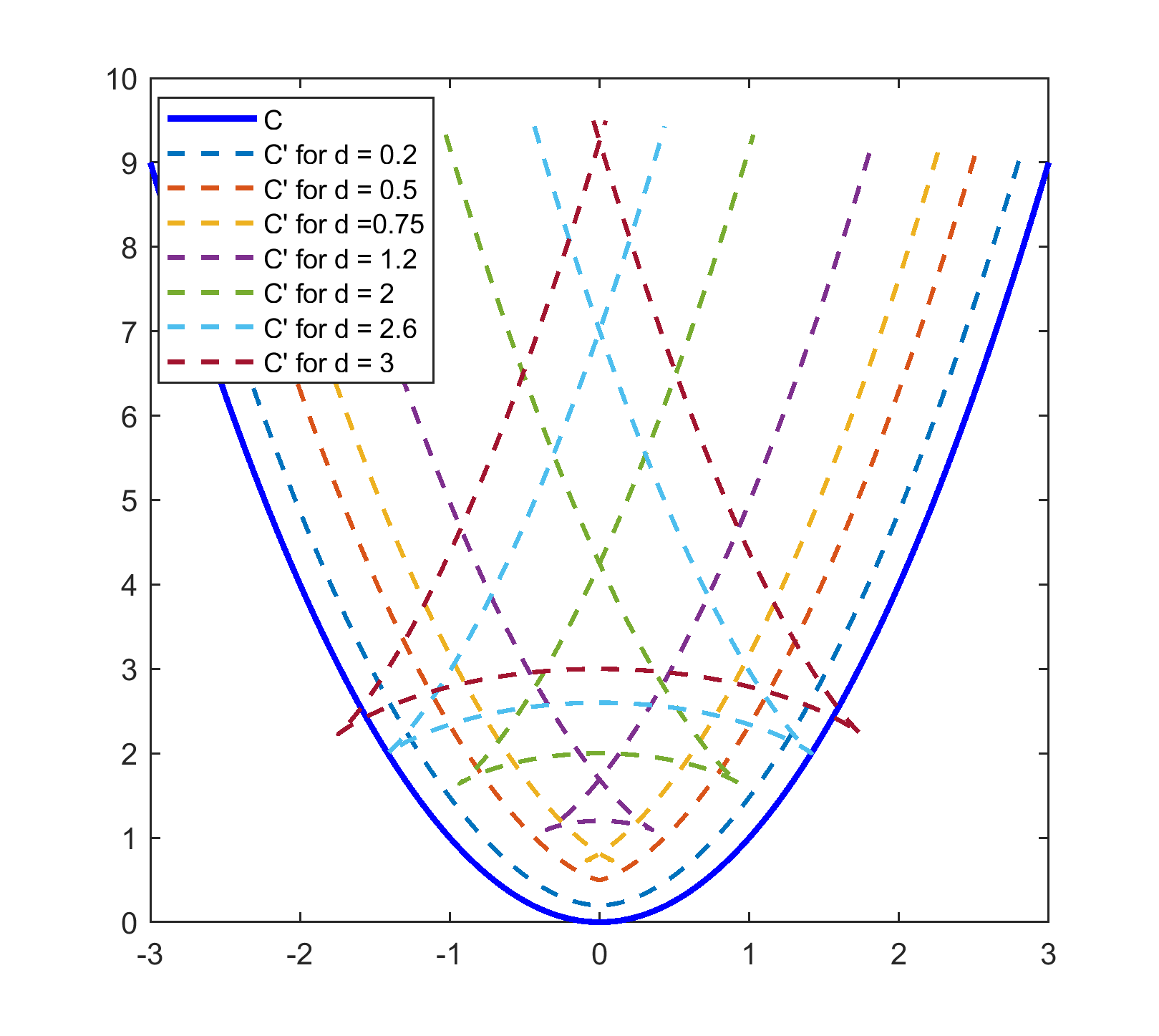}}

\subfigure[$f(x) = \exp{\sqrt{|x|}}$]{\label{fig:sub7}\includegraphics[width=0.45\linewidth]{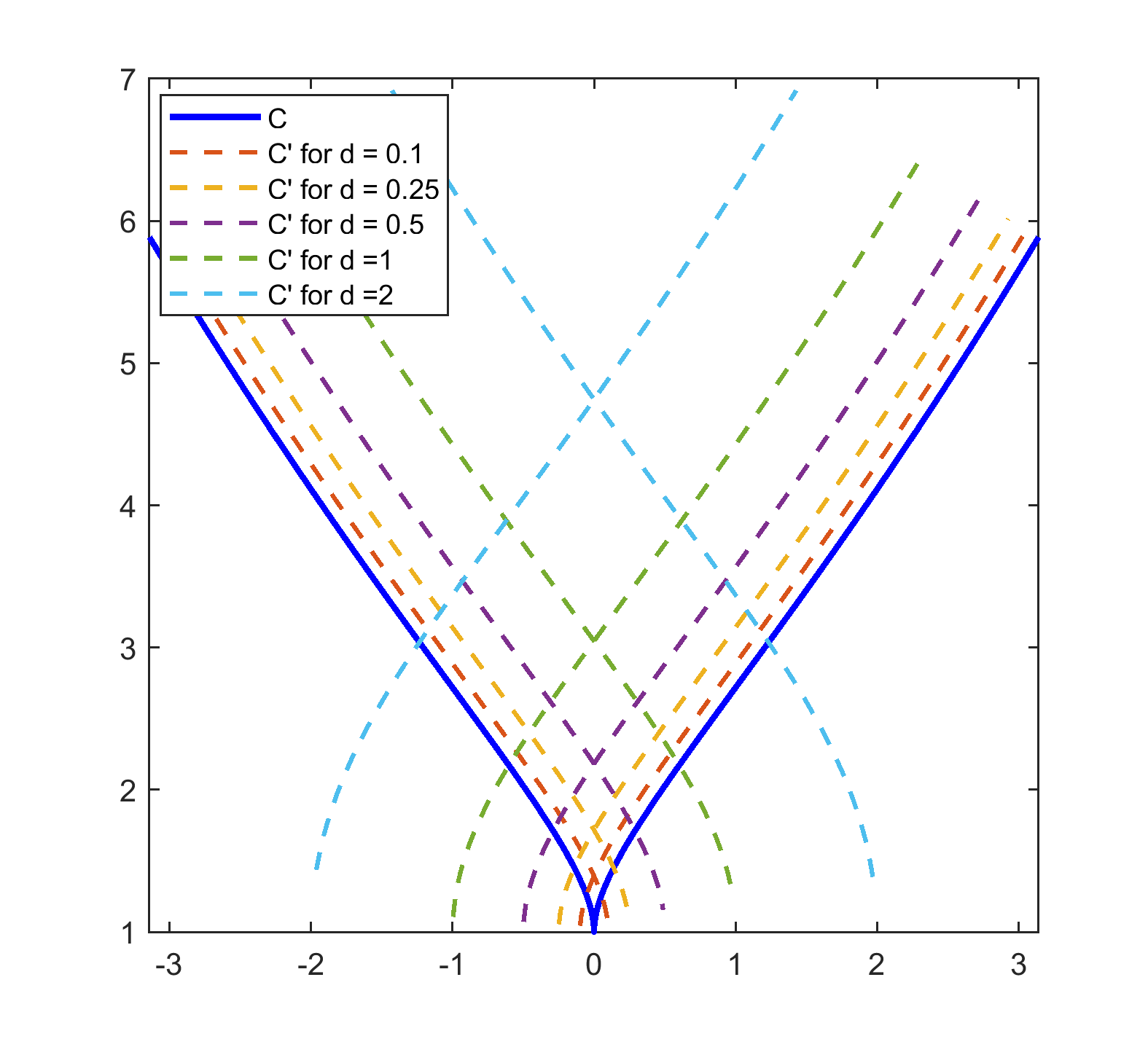}}
\hfill

\caption{\textbf{a.} As $d$ increases, $C'$ moves further away from $C$, thus the upper bound on $d$, $D = \infty$.
\textbf{b.} As $d$ increases, $C'$ moves further away from $C$. For lower values of $d$, $C'$s do not intersect $C$ and therefore are valid imaging curves. But for bigger values of $d$, they intersect $C$ and hence are invalid. Here the upper bound on $d$, $D \approx 2.6$.
\textbf{c.} Here $C$ is non-smooth and it is non-differentiable at $x = 0$. In this case, $C'$ generated for any $d>0$ is invalid, as it always intersects $C$, i.e. there are always points around $0$ which generate invalid imaging points. Also, as $d$ increases, the number of invalid imaging points increases.}
\label{fig:6}
\end{figure}
The imaging surfaces for some curves have been computed and plotted in \textsc{Matlab} as shown in figure \ref{fig:6}. The target curves($C$) are shown in blue along with the imaging curves($C'$) for different values of $d$. It is interesting to note that all non-smooth curves do not generate invalid imaging curves. For example, $f(x) = |mx|$ is non-differentiable at $x = 0$, but generates valid imaging curves for all $m \in (0,1]$.

\section{MODELLING ORTHOGRAPHIC IMAGING}
\label{sec:ortho_appx}
  
\subsection{Definition and Assumptions}
A practical approximation of orthography is to consider a very small($\epsilon$) angular \textit{field of view(FOV)} and the points on the surface within this $\epsilon$-FOV to be roughly orthographic and the boundary of the region consisting of such points is the \textit{orthographic boundary}. Here $\epsilon$ is a very small angle ($\sim \ 10^{\circ} - 20^{\circ}$). Also, for orthographic imaging of a surface point $P$, the imaging point must be along the surface normal at $P$ and at a height $d$. The region in a capture which lies within the orthographic boundary is defined as the \textit{$\epsilon$-Orthographic Image}.

\subsection{Circular Case}
\begin{figure}[H]
\centering
\subfigure[Imaging point at centre $O$]{\label{fig:sub11}\includegraphics[width=0.4\linewidth]{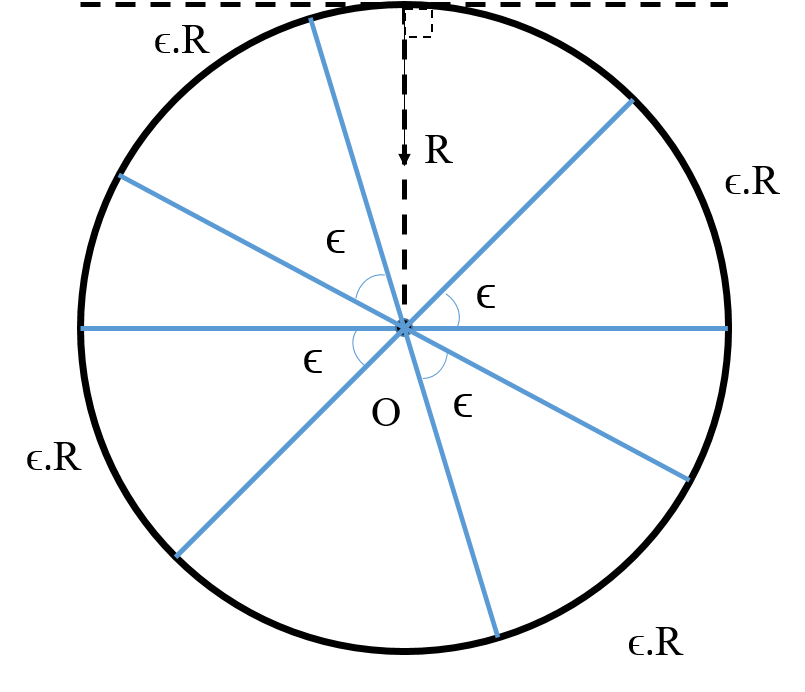}}
\hfill
\subfigure[Eccentric imaging point at $Q$ ]{\label{fig:sub12}\includegraphics[width=0.5\linewidth]{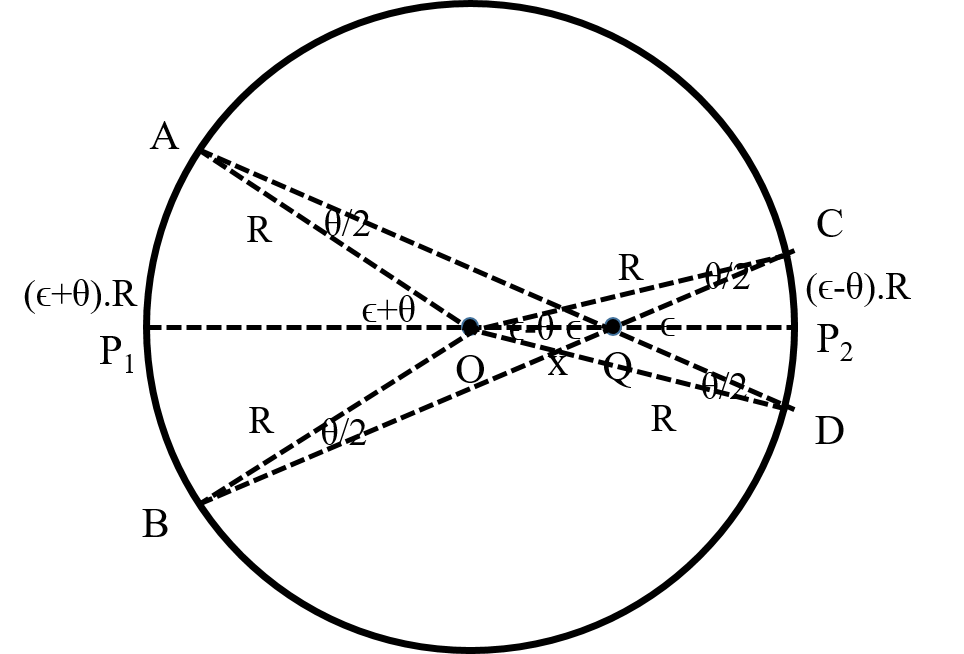}}
\caption{The case of circle as the target curve.}
\label{fig:7}
\vspace{-3mm}
\end{figure}

For a circle $C$ (Fig. \ref{fig:7}a), that radius to a point on the circumference is always orthogonal to the tangent at that point. Consequently, the centre $O$ of the circle satisfies the properties of a valid imaging point for any point $P$ on the circle. So the imaging point at $O$ can be used to capture a length of $2 \pi R$ or the entire circumference as shown in figure below. The total number of captures required to cover the entire circle is $\ceil{\frac{2\pi}{\epsilon}}$. Here imaging height is $d = R$.

For an imaging point located at an eccentric point $Q$ (Fig. \ref{fig:7}b) at distance $x$ from $O$, the normal from only two diagonally opposite points on the circumference passes through it. In this case the total length of $C$ that can be imaged can be proved to be $2\epsilon R$.

\subsection{Derivation of $\epsilon$-Orthography}
\subsubsection{Bounds for Curves}
\begin{figure}[H]
  \centering
  \includegraphics[width=0.5\linewidth]{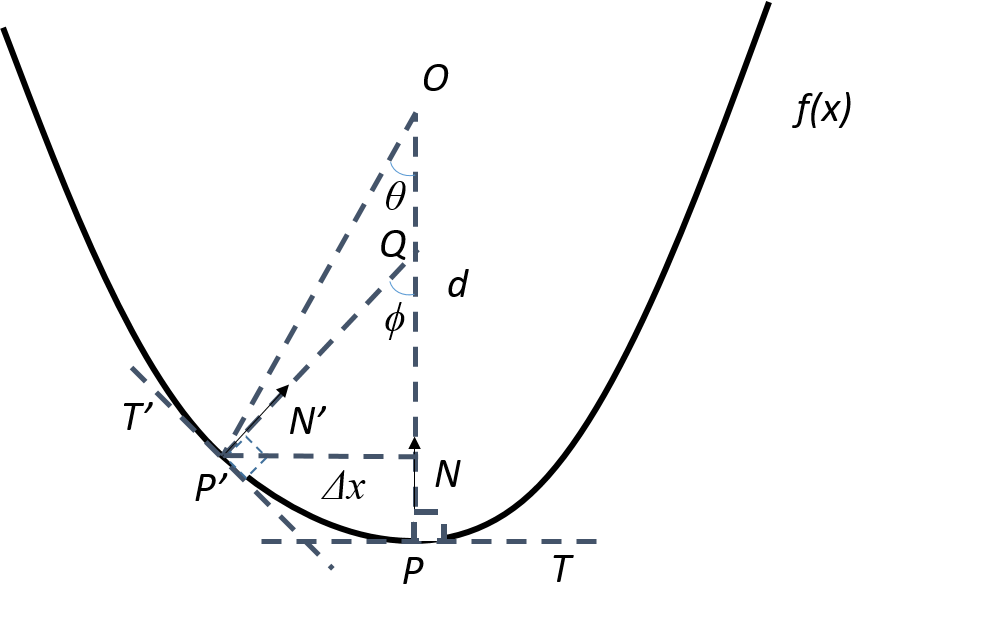}
  \caption{Illustrating the variables for derivation.}
  \label{fig:8}
\end{figure}

Let us consider a curve $C$ (Fig.\ref{fig:8}) given by a univariate function $f(x)$. The tangent($\Vec{T}$) and normal($\Vec{N}$) vectors at point $P(x,f(x))$ are given as
\begin{equation}
    \Vec{T}(x) =  \left[ \begin{array}{c}{1}\\{f'(x)}
    \end{array}\right] \ \ \ \ 
    \Vec{N}(x) = \left[ \begin{array}{c}{-f'(x)}\\{1}
    \end{array}\right].
\end{equation}

Let point $P'(x',f(x'))$ be situated at a small distance $\Delta x$ to the left of $x$. Let $p = f'(x)$ and $p' = f'(x'). $If $\Delta x$ is very small then $f(x')$ and $f'(x')$ can be approximated as
\begin{equation}
\begin{aligned}
    f(x') & = f(x - \Delta x) \approx f(x) - \Delta x\cdot f'(x) \\
    f'(x') & = f'(x - \Delta x) \approx f'(x) - \Delta x\cdot f''(x),
\end{aligned}
\end{equation}
therefore,
\begin{equation}
\begin{aligned}
    p' & = p + \Delta p \\
    \Delta p & \approx - \Delta x\cdot f''(x)
\end{aligned}
\end{equation}

Tangent $\Vec{T'}$ and normal $\Vec{N'}$ are constructed at $P'$. The normals $\Vec{N}$ and $\Vec{N'}$ intersect at $Q$ at an angle $\phi$. Therefore,
\begin{equation}\label{eq15}
    \begin{aligned}
    cos(\phi) & = \frac{\Vec{N}\cdot\Vec{N'}}{|\Vec{N}|\cdot|\Vec{N'}|} \\
              & = \frac{1}{\sqrt{p^{2} + 1}\sqrt{p'^{2} + 1}}\cdot\left[ \begin{array}{c}{-p}\\{1}
    \end{array}\right] \cdot \left[ \begin{array}{c}{-p'}\\{1}
    \end{array}\right] \\
              & = \frac{pp' + 1}{\sqrt{p^{2} + 1}\sqrt{p'^{2} + 1}}. 
    \end{aligned}
\end{equation}
So, 
\begin{equation}\label{eq16}
    \phi = cos^{-1}\Big(\frac{pp' + 1}{\sqrt{p^{2} + 1}\sqrt{p'^{2} + 1}}\Big).
\end{equation}
Also, if the line joining $P'$ and the imaging point $O$ intersect $\overline{OP}$ at angle $\theta$, as $\Delta x$ is much smaller compared to $d$,
\begin{equation}\label{eq17}
\begin{aligned}
    & tan(\theta)  = \frac{\Delta x}{d} \\
    or, \ & \theta  = tan^{-1}\big(\frac{\Delta x}{d}\big).
\end{aligned}
\end{equation}
$\epsilon$-orthographic bounds are dependent on both the FOV and the curvature at the concerned point.  For a point $P'$ on $C$ to lie within the $\epsilon$-orthographic region for capturing point $O$ at a height $d$ from the point $P$, it must satisfy-
\textbf{1.} $\theta \leq \epsilon$, and \textbf{2.} $\phi \leq \epsilon$.

\textit{Condition 1} is necessary so that the point $P'$ lies within the $\epsilon$-FOV. \textit{Condition 2} is required because as curvature of $C$ increases around $P$, although a point close to it may remain within the $\epsilon$-FOV bound, the high curvature causes very small region around $P$ to be approximately orthographic. With reference to Fig. \ref{fig:8}, if curvature at $P$ increases, $OP$ and $OP'$ may differ very much and then $P$ and $P'$ cannot be considered in the same orthographic region.

\subsubsection{Boundary for Surfaces}
The derivation is very similar to that for the curves. A surface $S$ is given by a bi-variate function, $z = f(x,y)$. Given a central point $P(x,y,z)$ on the surface, an imaging height of $d$ and useful FOV $\epsilon$, the goal is to find the orthographic boundary surrounding $P$, or in other words, the area around $P$ which can be considered as an approximate orthographic image. 

From Eqs. \ref{eq2} and \ref{eq3}, the surface normal at point $P(x,y,z)$ is $[p, q, -1]$, where $p = \frac{\partial f(x,y)}{\partial x}$ and $q = \frac{\partial f(x,y)}{\partial y}$. The Hessian matrix of $f(x,y)$ is 
\begin{equation}
    H=\left[\begin{array}{cc}{\frac{\partial^{2} f(x, y)}{\partial x^{2}}} & {\frac{\partial^{2} f(x, y)}{\partial x \partial y}} \\ {\frac{\partial^{2} f(x, y)}{\partial y \partial x}} & {\frac{\partial^{2} f(x, y)}{\partial y^{2}}}\end{array}\right].
\end{equation}
Let us take a point $P'(x',y',z')$ very close to $P$ such that
\begin{equation}
    \begin{aligned}
    x' & = x + \Delta x \\
    y' & = y + \Delta y.
    \end{aligned}
\end{equation}
As $\Delta x$ and $\Delta y$ are very small quantities, the change in the  surface normal vector is also small. So the surface normal at $P'$, $\Vec{N'} =  [p', q', -1]$ and it can be approximated as
\begin{equation}
    \begin{aligned}
    & \left[\begin{array}{cc}{\Delta p}\\{\Delta q}\end{array}\right] =  H\cdot \left[\begin{array}{cc}{\Delta x}\\{\Delta y}\end{array}\right] \\
    & p' = p + \Delta p \\
    & q' = q + \Delta q \\
    \end{aligned}
\end{equation}
Similarly as Eqs. \ref{eq15} and \ref{eq17}, $\phi$ and $\theta$ are calculated as 
\begin{equation}
    \begin{aligned}
    cos(\phi) & = \hat{N}.\hat{N'} \\
              & = \frac{\Vec{N}\cdot\Vec{N'}}{|\Vec{N}||\Vec{N'}|} \\
              & = \frac{1}{\sqrt{p^{2} + q^{2} + 1}\sqrt{p'^{2} + q'^{2} + 1}}\left[ \begin{array}{c}{p}\\{q}\\{-1}
    \end{array}\right] \cdot \left[ \begin{array}{c}{p'}\\{q'}\\{-1}
    \end{array}\right] \\
              & = \frac{pp' + qq' + 1}{\sqrt{p^{2} + q^{2} + 1}\sqrt{p'^{2} + q'^{2} + 1}}. 
    \end{aligned}
\end{equation}

So, 
\begin{equation}\label{eq22}
    \phi = cos^{-1}\Big(\frac{pp' + qq' + 1}{\sqrt{p^{2} + q^{2} + 1}\sqrt{p'^{2} + q'^{2} + 1}}\Big)
\end{equation}

and
\begin{equation}\label{eq23}
\begin{aligned}
    & tan(\theta)  = \frac{\sqrt{\Delta x^{2} + \Delta y^{2}}}{d} \\
    or, \ & \theta  = tan^{-1}\Big(\frac{\sqrt{\Delta x^{2} + \Delta y^{2}}}{d}\Big).
\end{aligned}
\end{equation}

Now, if $P'$ belongs to the orthographic region around point $P$ for an imaging height $d$, then both $\theta \leq \epsilon$ and $\phi \leq \epsilon$.
In all the aforementioned derivations, the gradient components at $P'$ have been calculated by approximations for fast computation. Where computational capability is not an issue, the actual gradients can be calculated by differentiating the function $f(x,y)$.
\section{IMPLEMENTATION}\label{sec:implement}
\label{sec:implement}
\subsection{Algorithms}

The algorithm for computing the orthographic bounds for a smooth curve $f(x)$ at a central point $P_{0}(x_{0},f(x_{0}))$, for $\epsilon$ angular FOV, resolution $dx$ and imaging height $d$ is stated.

\begin{algorithm}[hbt!]
\scriptsize
\caption{Finding Orthographic Bounds of a Curve }\label{alg:ortholeft}
\begin{algorithmic}[1]
\Procedure{Left Orthographic Bound}{$x_{0},\epsilon, d, dx$}
    \State Find $p_{0} = f'(x_{0})$
    \State Set $x = x_{0} - dx$
    \State Find $p = f'(x)$ and calculate $\phi$ using Eq.\ref{eq16}
    \State Set $\Delta x = |x - x_{0}|$
    \State Calculate $\theta$ using Eq.\ref{eq17}
    \While{$\phi \leq \epsilon$ and $\theta \leq \epsilon$}
        \State Set $x = x - dx$
        \State Find $p = f'(x)$ and calculate $\phi'$ using Eq.\ref{eq16}
        \State $\phi = \phi'$'
        \State Set $\Delta x = |x - x_{0}|$
        \State Calculate $\theta'$ using Eq.\ref{eq17}
        \State $\theta = \theta'$
    \EndWhile
    \State \textbf{return} $x = x + dx$
\EndProcedure
\Procedure{RightOrthographic Bound}{$x_{0},\epsilon, d, dx$}
    \State Find $p_{0} = f'(x_{0})$
    \State Set $x = x_{0} + dx$
    \State Find $p = f'(x)$ and calculate $\phi$ using Eq.\ref{eq16}
    \State Set $\Delta x = |x - x_{0}|$
    \State Calculate $\theta$ using Eq.\ref{eq17}
    \While{$\phi \leq \epsilon$ and $\theta \leq \epsilon$}
        \State Set $x = x + dx$
        \State Find $p = f'(x)$ and calculate $\phi'$ using Eq.\ref{eq16}
        \State $\phi = \phi'$'
        \State Set $\Delta x = |x - x_{0}|$
        \State Calculate $\theta'$ using Eq.\ref{eq17}
        \State $\theta = \theta'$
    \EndWhile
    \State \textbf{return} $x = x - dx$
\EndProcedure

\end{algorithmic}
\end{algorithm}
\vspace{-3mm}
\begin{figure}[hbt!]
\centering
\subfigure[Convex part]{\label{fig:sub11}\includegraphics[width=0.48\linewidth]{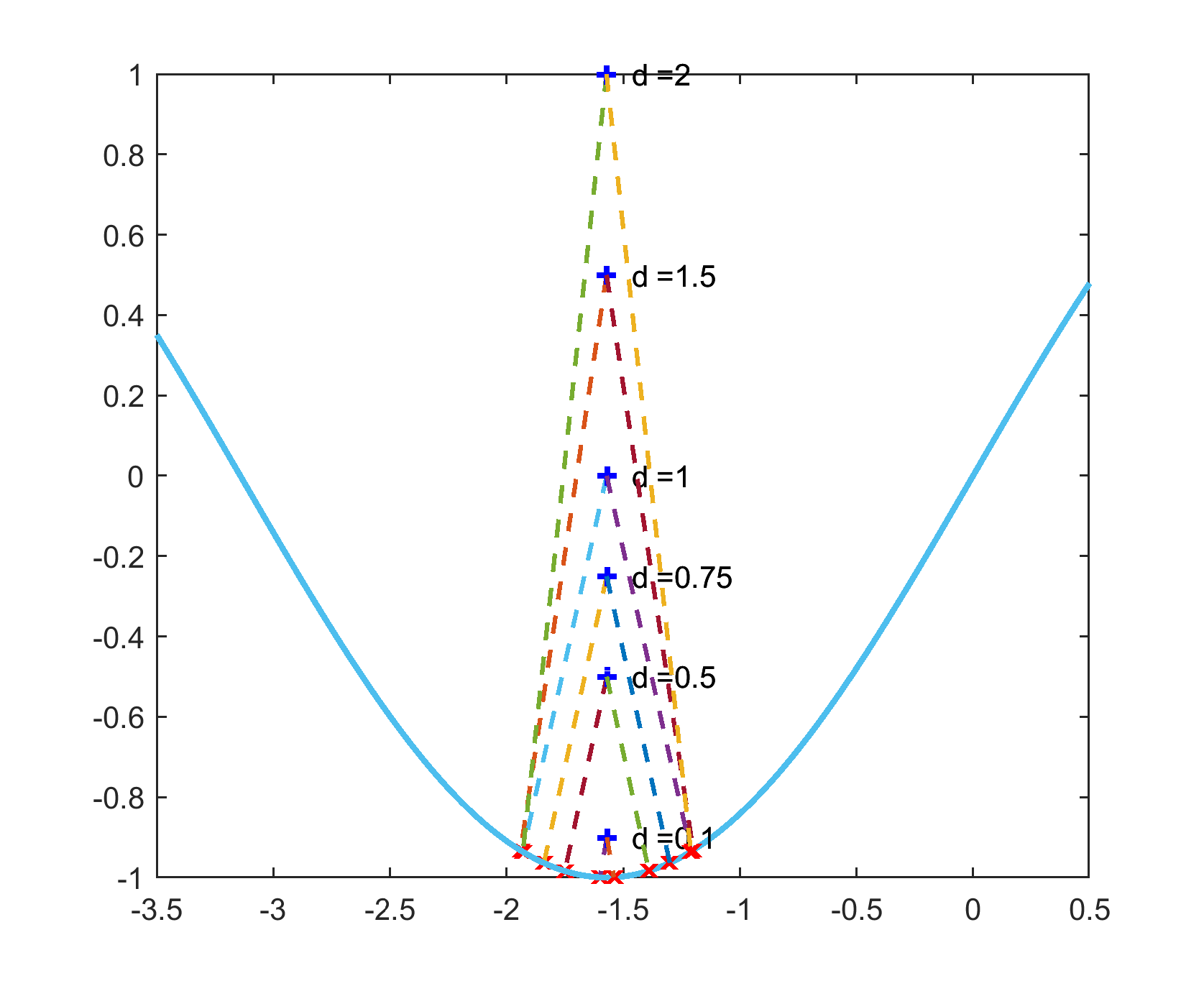}}
\hfill
\subfigure[Concave part]{\label{fig:sub12}\includegraphics[width=0.48\linewidth]{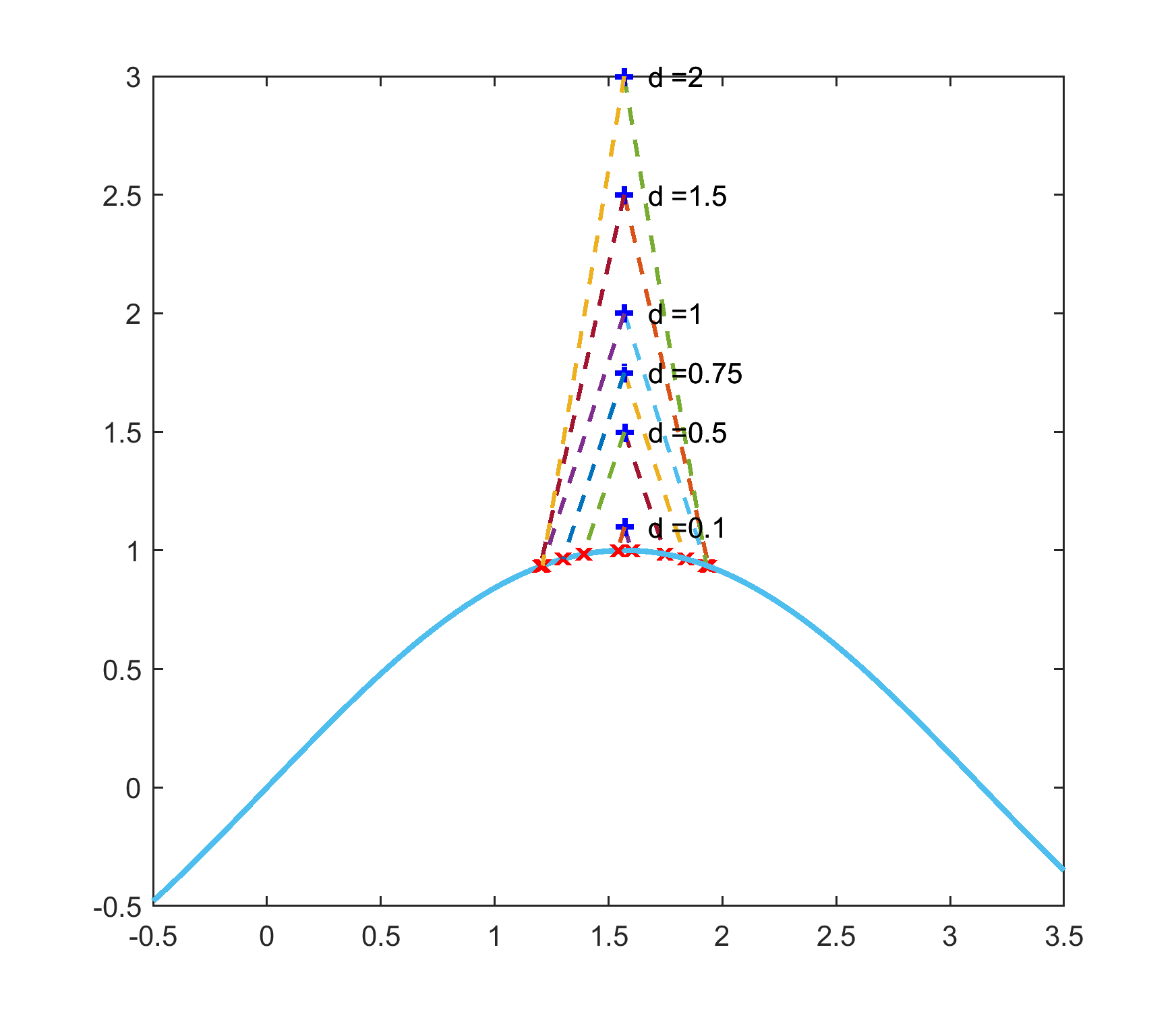}}
\caption{The orthographic bounds for convex and concave parts of a sine curve for increasing $d$.}
\label{fig:9}
\vspace{-3mm}
\end{figure}

Algorithm \ref{alg:ortholeft} has been implemented in \textsc{Matlab} and the bounds have been calculated and plotted for a convex curve and a concave curve as shown in Fig. \ref{fig:9}. It is to be noted that although $d$ keeps increasing, the bounds do not spread after a point. If the bounds were a function of $\epsilon$ only, then with increase in $d$, they would have spread apart indefinitely which is not a true characteristic of orthography. This shows that $\epsilon$-orthography not only depends on the FOV but also the curvature.

The $\epsilon$-orthographic boundary can be numerically calculated using Algorithm \ref{alg:orthoboundary} for a surface $S$ ($z = f(x,y)$) at a central point $P_{0}(x_{0},y_{0},z_{0})$, for $\epsilon$ angular FOV, resolutions $dx$ and $dy$, and imaging height $d$.
\begin{algorithm}[hbt!]
\scriptsize
\caption{Finding Orthographic Boundary of a Surface }\label{alg:orthoboundary}
\begin{algorithmic}[1]
\State Find surface normal components $p_{0}$ and $q_{0}$ at $P_{0}$.
\State Create an empty point set $P$ for storing the eligible points inside the orthographic region.
\State Append $P_{0}$ to $P$
\State Compute the number of $x$ or $y$ co-ordinates in the grid. $n_{x} = (r_{H} - r{L})/dx = R_{x}/dx$ and $n_{y} = R_{y}/dy$.
\State Set $s = max(n_{x},n_{y})$
\State Set $buff = 0$
\For{$n = 1:s$}
    \State Vector $out = PairGen(n)$
    \State Set $count = 0$
    \For{$j = 1:length(out)$}
        \State $x_{1} = x_{0} + dx\cdot out(1,j)$
        \State $y_{1} = y_{0} + dy\cdot out(2,j)$ 
        \State Compute surface normal vector at $P_{1}(x_{1},y_{1},z_{1})$
        \State Calculate $\phi$ using Eq.\ref{eq22}
        \State Calculate $\theta$ using  Eq.\ref{eq23}
        \If{$\theta \leq \epsilon \ \ and \ \ \phi \leq \epsilon$}
            \State Append $P_{1}(x_{1},y_{1},z_{1})$ to 
            \State $count = count + 1$
        \EndIf
    \EndFor
    \If{$count = 0$}
        \State $buff = buff + 1$
    \EndIf
    \If{$buff > 3$}
        \State Break  from loop
    \EndIf
\EndFor
\end{algorithmic}
\end{algorithm}
\vspace{-3mm}
\begin{figure}[htb!]
\centering
\subfigure[$d = 2, \ (x_{0},y_{0}) = (0,0)$]{\label{fig:sub5}\includegraphics[width=0.45\linewidth]{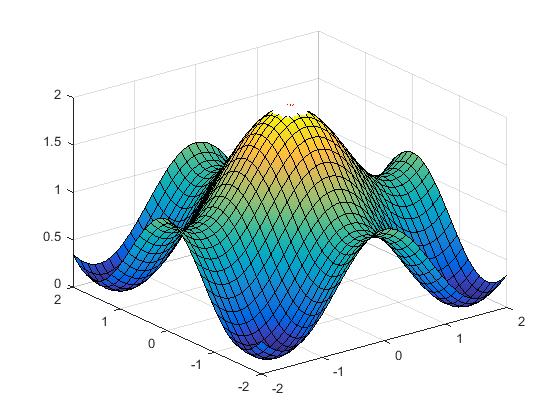}}
\hfill
\subfigure[$d = 2, \ (x_{0},y_{0}) = (0,0)$]{\label{fig:sub6}\includegraphics[width=0.45\linewidth]{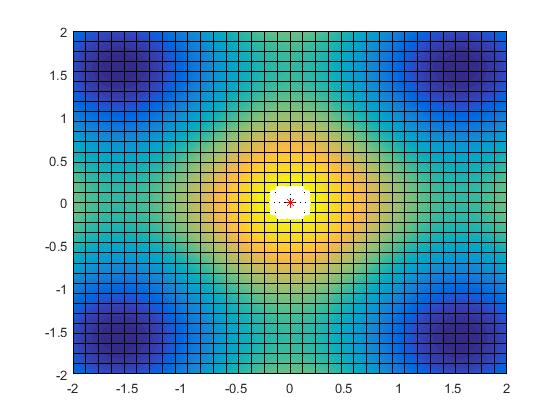}}

\subfigure[$d = 2, \ (x_{0},y_{0}) = (0,-1)$]{\label{fig:sub11}\includegraphics[width=0.45\linewidth]{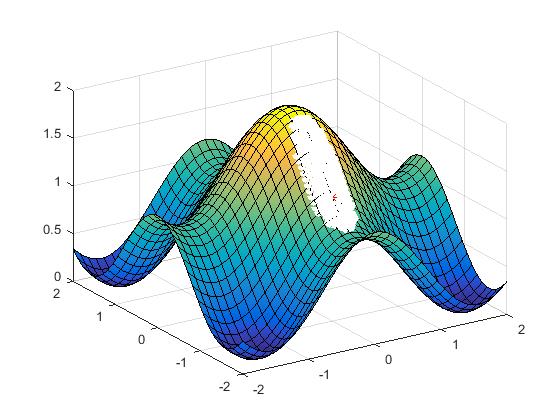}}
\hfill
\subfigure[$d = 2, \ (x_{0},y_{0}) = (0,-1)$]{\label{fig:sub12}\includegraphics[width=0.45\linewidth]{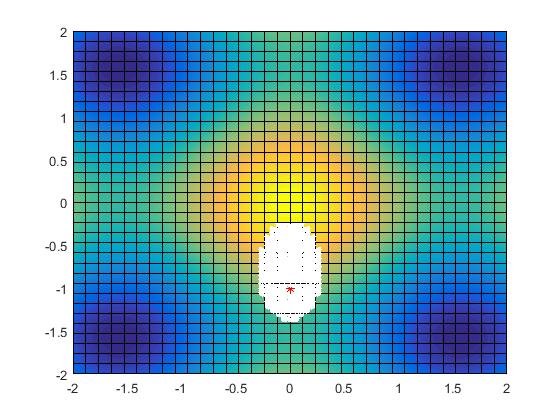}}

\subfigure[$d = 2, \ (x_{0},y_{0}) = (-1,-1)$]{\label{fig:sub7}\includegraphics[width=0.45\linewidth]{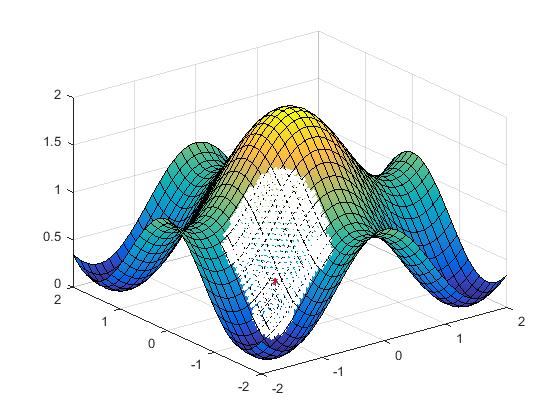}}
\hfill
\subfigure[$d = 2, \ (x_{0},y_{0}) = (-1,-1)$]{\label{fig:sub8}\includegraphics[width=0.45\linewidth]{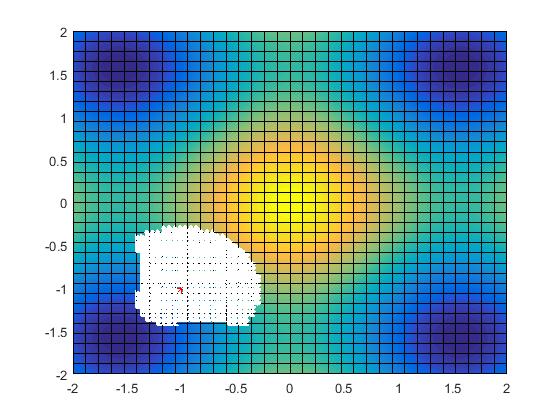}}

\caption{Orthographic regions drawn on curve $f(x,y) = cos^{2}(x) + cos^{2}(y)$ shown in white. The figures on right show the boundary shape. The central point $(x_{0},y_{0})$ is plotted in red. (Here $\epsilon = 10^{\circ}$)}
\label{fig:10}
\vspace{-3mm}
\end{figure}
The function \textit{PairGen} generates a vector of all pairs of integers $n_{1}$ and $n_{2}$ such that $|n_{1}| + |n_{2}| = n$, i.e. all co-ordinates located at absolute distance $n$. Algorithm \ref{alg:orthoboundary} has been implemented in \textsc{Matlab} to generate $\epsilon$-orthographic regions on smooth curves (Fig. \ref{fig:10}). This algorithm is also valid for non-smooth curves, for which numerical gradients can be calculated at non-differentiable points.
\vspace{-2mm}
\subsection{Special Surfaces}
\textbf{\textit{Conjecture:}} Points on surfaces of constant Gaussian curvature \cite{gauss_curv} form $\epsilon$-orthographic regions of same area for constant imaging height $d$. The upper bound on $d$ depends on the nature (parameters) of such surfaces.

Surfaces of constant curvatures can be classified into the following three classes-

{\textit{1) \textbf{Zero Curvature Surfaces: }}}
A surface with Gaussian curvature($\kappa$) equal to zero at all points is a plane. For a plane, which is inherently orthographic, the calculated region is thus of same shape as the \textit{FOV}. As the \textit{FOV} is considered circular in all our calculations, the orthographic region is thus circular for a planar surface, the radius of which depends on the imaging height as given by Eq.\ref{eq24}. Thus the problem of finding optimal capture points is reduced to a \textit{Circle Packing} problem.

\begin{figure}[hbt!]
\centering
\subfigure[3D view of the surface]{\label{fig:sub64a}\includegraphics[width=0.35\linewidth]{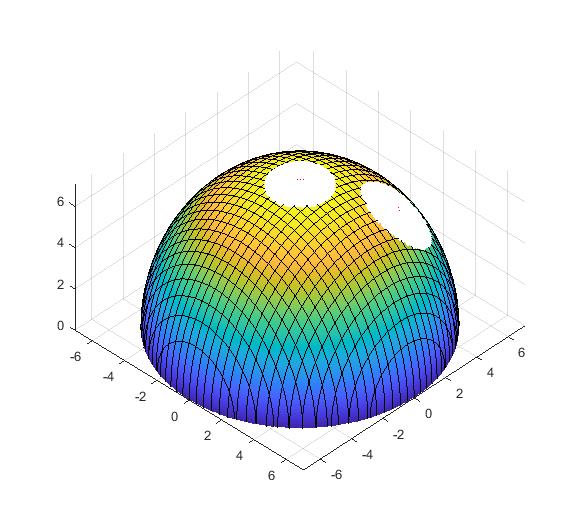}}
\hfill
\subfigure[2D top view of the surface]{\label{fig:sub64b}\includegraphics[width=0.35\linewidth]{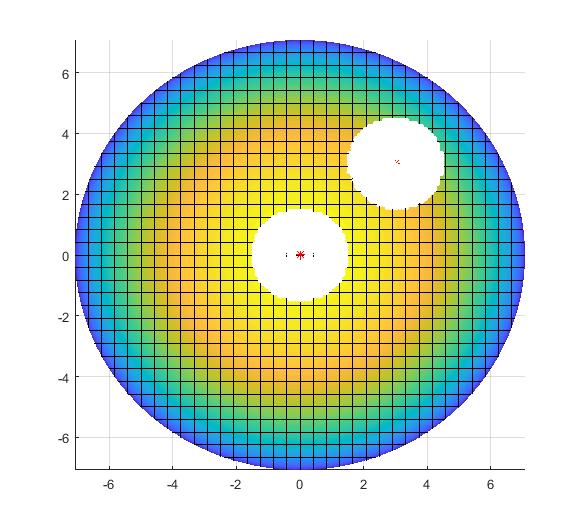}}
\vfill
\subfigure[3D view of the surface]{\label{fig:sub65a}\includegraphics[width=0.35\linewidth]{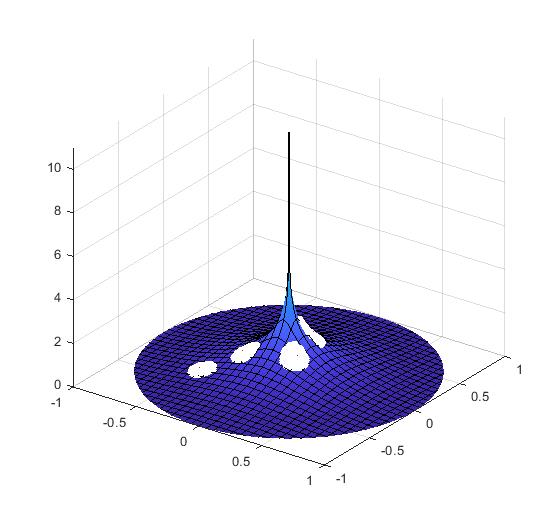}}
\hfill
\subfigure[2D top view of the surface]{\label{fig:sub65b}\includegraphics[width=0.35\linewidth]{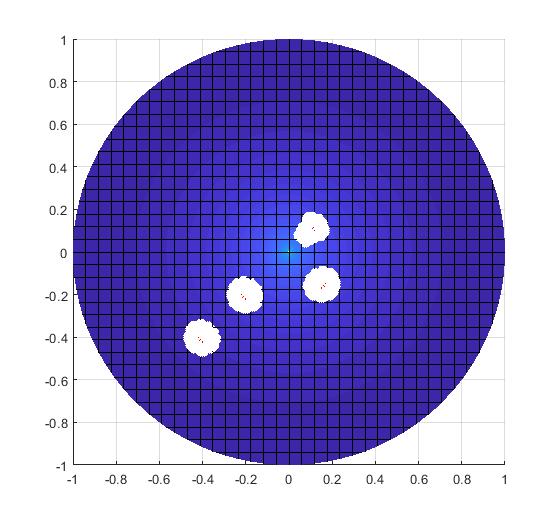}}

\caption{\textbf{a--b.} $\epsilon$-Orthographic regions plotted on a sphere- a surface of constant positive Gaussian curvature. \textbf{c--d.} $\epsilon$-Orthographic regions plotted on a pseudosphere- a surface of constant negative Gaussian curvature (large regions are plotted for demonstration).}
\label{fig:11}
\vspace{-5mm}
\end{figure}

{\textit{2) \textbf{Positive Curvature Surfaces: }}}
A surface with equal positive Gaussian curvature($\kappa$) at all points is sphere.
Using the definition of $\epsilon$-orthography, it can be shown that for a sphere, the orthographic regions are also circular and of constant radii, dependent on the imaging height $d$ (Fig.\ref{fig:11}a--b). This is due to the fact that the two principal curvatures ($\kappa_1$ and $\kappa_2$) at any point on the sphere are equal and constant. However, unlike a plane, sphere is not inherently orthographic but behaves like one. 

{\textit{3) \textbf{Negative Curvature Surfaces: }}}
A surface with equal negative Gaussian curvature($\kappa$) at all points is a pseudosphere \cite{pseudo1999}. Unlike the other two cases, for a pseudosphere, the orthographic boundaries are not circular, and the limit on imaging height $d$ is dependent on the radius $a$ (Fig.\ref{fig:11}c--d). Among the two principal curvatures ($\kappa_1$ and $\kappa_2$) calculated at any point on the surface, one is positive and the other is negative. As we move along the surface, from the flat region to the narrow region, the magnitude of the positive principal curvature increases and the negative principal decreases such that the product of the two (Gaussian curvature) remains the same. As a consequence of this property, it has been empirically observed that the size of the $\epsilon$-orthographic regions remain the same, although the shape may vary (figure \ref{fig:11}c--d).
\vspace{-2mm}
\subsection{Limitations}\label{subsec:ref}
The calculation of orthographic regions is computationally expensive, specially for higher resolutions. Consequently, finding the overlap between two regions is also expensive. Also, the entire orthographic region needs to be calculated to find the boundary. 
For non-smooth surfaces, gradients cannot be calculated at non-differentiable points and 
for natural surfaces, with fast varying curvatures, orthographic boundaries are difficult to calculate.
Moreover, the exact boundaries cannot be used to calculate optimal capture points, where the boundaries need to be computed at multiple points simultaneously, which is repetitive and slow. 

\section{APPROXIMATION OF ORTHOGRAPHIC BOUNDARY}\label{sec:appx_boundary}
\subsection{Approaches}
As pointed out in the limitations (\ref{subsec:ref}), calculating the exact orthographic region and hence the boundary is not practical because it is computationally expensive and thus time consuming. However, instead of considering exact boundaries, they can be approximated to some regular shapes for faster boundary computation as well as calculating overlap between regions as explored in the following approaches.



\begin{figure}[hbt!]
\centering

\subfigure[Polygonal]{\label{fig:sub13}\includegraphics[width=0.32\linewidth]{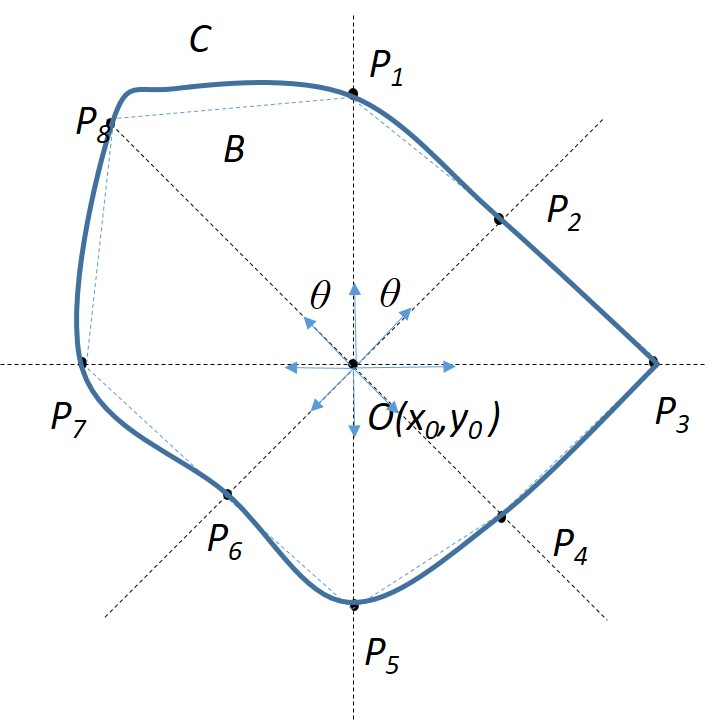}}
\hfill
\subfigure[Elliptical]{\label{fig:sub14}\includegraphics[width=0.32\linewidth]{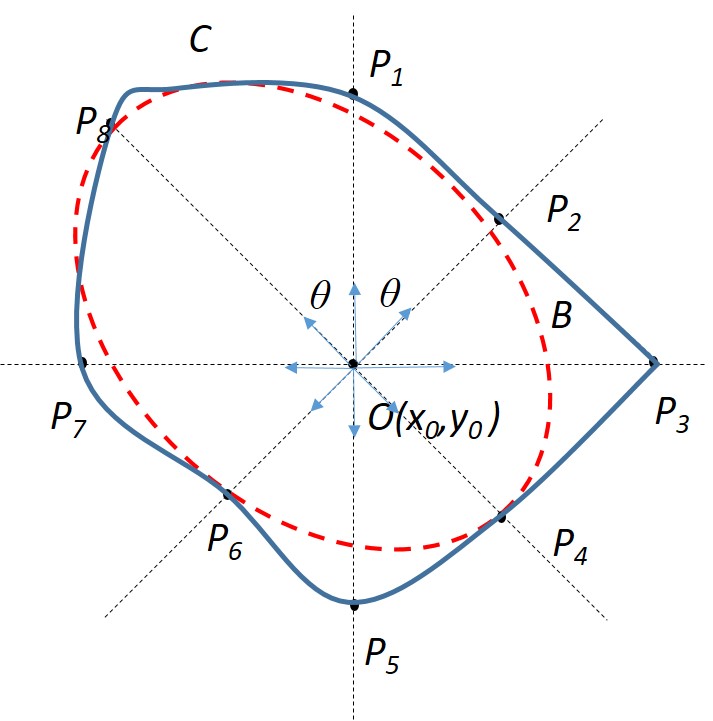}}
\hfill
\subfigure[Circular-I]{\label{fig:sub15}\includegraphics[width=0.32\linewidth]{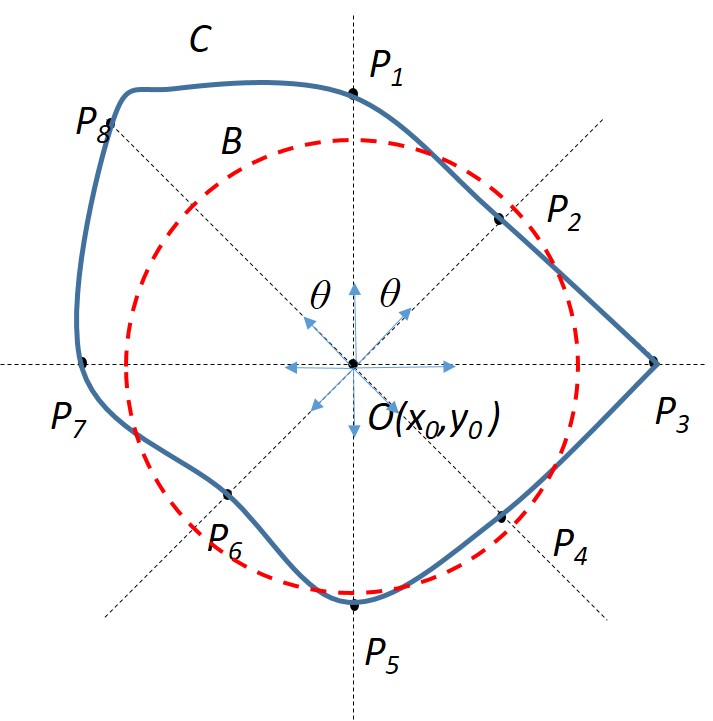}}

\caption{\textbf{a.} Boundary points detected for $N = 8$. Here $\theta = 45^{\circ}$. Actual boundary $C$ shown in bold. 
\textbf{b.} Actual boundary $C$ and approximated elliptical boundary $B$.
\textbf{c.} Actual boundary $C$ shown in blue and approximated circular boundary $B$ shown in red.}
\label{fig:13}
\vspace{-5mm}
\end{figure}
{\textit{1) \textbf{Polygonal Approximation: }}}
In this approach, $N$ points are calculated in $N$ different directions from the central point. By setting $\Delta x$ and $\Delta y$ in Eq.\ref{eq22} according to the direction of calculation, and by calculating $\theta$ and $\phi$ using Eqs.\ref{eq22} and \ref{eq23}, and checking at each step to see whether they maintain the $\epsilon$ constraint, the boundary point in the concerned direction can be computed. The directions in which the boundary points must be calculated, should be at equal angles to each other at the central point $(x_{0}, y_{0})$, i.e. the directions should be at an angle $\theta = \frac{360^{\circ}}{N}$ from each other. The $N$ boundary points are joined to form the $N$-polygonal boundary.

Fig.\ref{fig:sub13} shows an example of polygonal approximation of the boundary. Here the boundary points $P_{i}$'s are evaluated in 8 directions centered at $O(x_{0},y_{0})$. Obviously, larger the number of directions taken, better will be the approximate boundary. This approach may lead to both over-estimation and under-estimation of boundary depending on the convexity of the boundary curve.

{\textit{2) \textbf{Elliptical Approximation: }}}
This is an extension or further approximation of the \textit{polygonal approximation} approach but here only even sided polygons are considered.
Boundary points are calculated in $N$ different equiangularly spaced directions. Hence, we get $N$ boundary points $P_{i}, \ \ i = 1,2,...,N$. 
Now, the distances between the diagonally opposite boundary points is calculated, and thus we have $N/2$ diagonals ($d_{i}$).
The maximum and minimum diagonals are considered, $d_{max} = max(d_{i})$ and $d_{min} = min(d_{i})$.
The boundary is approximated as an ellipse with the major axis as $d_{max}$ and the minor axis as $d_{min}$ and the major axis is aligned along the longest diagonal.

In Fig.\ref{fig:sub14}, the maximum length diagonal is $\overline{P_{4}P_{8}}$ and the minimum length diagonal is $\overline{P_{2}P_{6}}$. The major axis of the constructed ellipse($B$) is $\overline{P_{4}P_{8}}$ and the minor axis is of same length as  $\overline{P_{2}P_{6}}$. It is to be noted that central point $O(x_{0},y_{0})$ is not the centre of the ellipse.

{\textit{3) \textbf{Circular Approximation--I: }}}
This is a further simplification of the \textit{elliptical approximation}. In this case, the orthographic boundary is approximated as a circle.
Boundary points are calculated in $N$ different equiangularly spaced directions as discussed in \textit{polygonal} case. Hence, we get $N$ boundary points $P_{i}, \ \ i = 1,2,...,N$. 
Now, the distances of the boundary points from the central point $(x_{0},y_{0})$ is calculated, and thus we have $N$ distances ($d_{i}$).
The average of all the distance lengths is calculated, $d_{avg} = \Sigma d_{i}$.The boundary is approximated as a circle centered at $(x_0,y_0)$ and of radius $R = d_{avg}$.
In the illustrated figure \ref{fig:sub15}, $d_{i} = \overline{OP_i}$, and the average of all 8 $\overline{OP_i}$'s is calculated. The boundary circle $B$ is constructed with centre at $O(x_0,y_0)$ and radius equal to the average of $OP_i$'s. 

{\textit{4) \textbf{Circular Approximation--II: }}}
This approach is the simplest and most intuitive approach among the ones discussed. It is based on the observation that on a smooth surface, the orthographic regions are small at places of high \textit{Absolute Gaussian Curvature} \cite{gauss_curv} and relatively larger at places where it is low.

If the orthographic region or boundary is estimated by a circle, an inverse relation between the radius of the boundary and the curvature of the central point can be formulated. Also, for a planar surface or a zero curvature surface, the orthographic region is circular with radius
\begin{equation}\label{eq24}
    R = d\cdot tan(\epsilon),
\end{equation}
where $d$ is the imaging distance and $\epsilon$ is the useful FOV as discussed in the derivation of $\epsilon$-orthography. For any point on a non-planar surface having an absolute curvature $|K| \geq 0$, the boundary will shrink from this circle. Therefore, if the region boundary is approximated by a circle of radius $r$, $r \leq R$.

Now, let us consider a surface $S$ and its Gaussian curvature ($K$). Given the bounds of the surface, the maximum absolute curvature is calculated.
\begin{equation}
    K_{max} = \max_{x,y}|K(x,y)|, \ \ \ \ (x,y) \ \in \ Bound(S).
\end{equation}
Fix a ratio ($m$) between the largest radius possible $(r_{max} = R)$ for the points of least absolute curvature and the least radius possible for the point having curvature $K_{max}$. So, $m = \frac{r_{max}}{r_{min}}$. Therefore, the radius of the approximated circular boundary can be expressed as a function of point $(x,y)$ as
\begin{equation}
    r(x,y) = R - \frac{|K(x,y)|}{K_{max}}\cdot R \cdot(1 - \frac{1}{m}).
\end{equation}
The value of $m$ can be tuned by experimental observations.

\subsection{Comparison and Analysis}
The four different approaches for approximating the orthographic boundary can be compared in terms of accuracy of approximation and computation time. For \textit{Approach 1} (\textit{Polygonal Approximation}), \textit{2} (\textit{Elliptical Approximation}) and \textit{3} (\textit{Circular Approximation - I}), the computational time depends on the number of directions $N$ or the number of boundary points used for approximation, which increases with increase in $N$. However in \textit{Approach 2}, the diagonals have to be compared and the equation of ellipse has to be calculated, which is the most time consuming among the four. For \textit{Approach 1} and \textit{2}, calculation of overlap among the regions is most complicated because of their polygonal and elliptical shape respectively, whereas, in \textit{Approach 3} and \textit{4}, it is much easier due to their circular shape. The accuracy of approximation decrease from \textit{1} to \textit{4}, polygonal being the most accurate and the second circular approximation being the crudest. However, \textit{Approach 4} is the most intuitive and easiest to compute and can be used for further optimization but has the highest error, specially for natural or fast-varying surfaces.

\section{CONCLUSION AND FUTURE WORK}\label{sec:conclusion}
Orthographic imaging is a crucial tool for terrain survey or terrain mapping. Although technological improvements have been made widely in the devices to capture visual or other sensor data of a surface, a proper and efficient algorithm for reconstructing the surface topography and creating an orthographic projection of the terrain is lacking. This paper aims to study and analyze this problem and propose novel methods for solving it. A technique for generating topographical surface from elevation maps has been proposed. A detailed study of imaging surfaces and bounds on imaging height has been presented. The effects of imaging height and angular field of view for capturing orthographic views have been formulated and analyzed in detail. A novel method for calculating orthographic boundaries have been proposed and demonstrated. Different methods of approximating the orthographic boundaries have been proposed and compared.

As a future extension of this study, better approximations of orthographic boundaries should be explored, and in case of which, how and by how much the results are affected must be studied. Methods for computing orthographic views by combining visual data with that from devices not capturing visual data can be explored and incorporated. Faster algorithms for real-time computation of orthographic boundaries should be explored for large-scale computation problems. The approximation can be used for computing the optimal number of capture points to cover the whole surface with minimum overlap among them.
\label{sec:conclusion}

\bibliographystyle{plain}
\bibliography{refs}

\begin{thebibliography}{1}

\bibitem{telelens2}
Telecentric lenses: {B}asic {I}nformation and {W}orking principles.
\newblock {\em Opto Engineering}, 2008.

\bibitem{imgparam}
Gregory Hollows.
\newblock Imaging {O}ptics {F}undamentals.
\newblock {\em Proceedings of the 17th {ACM} {SIGSPATIAL} international
  conference on advances in geographic information systems}, pages 1--16, 2014.

\bibitem{terrainmap2001}
Steven Scheding, Jeff Leal, Mark Bishop, and Salah Sukkarieh.
\newblock Terrain {M}apping in {R}eal-{T}ime: Sensors and {A}lgorithms.
\newblock {\em Geospatial Information and Agriculture}, 2001.

\bibitem{orthograph1}
Gary~S. Smith.
\newblock Digital {O}rthophotography and {GIS}.
\newblock {\em ESRI Conference}.

\bibitem{pseudo1999}
H~Steinhaus.
\newblock {\em Mathematical {S}napshots, $3^{rd}$ ed.}
\newblock New York: Dover, 1999.

\bibitem{gauss_curv}
Eric~W. Weisstein.
\newblock {\em Gaussian {C}urvature}.
\newblock
  \hyperlink{http://mathworld.wolfram.com/GaussianCurvature.html}{Mathworld}--
  A Wolfram Web Resource.

\end{thebibliography}

\end{document}